# Performance of n-in-p pixel detectors irradiated at fluences up to $5 \times 10^{15}$ $n_{eq}/cm^2$ for the future ATLAS upgrades

A. Macchiolo[a,*], C. Gallrapp[b], A. La Rosa[b,d], R. Nisius[a], H. Pernegger[b], R.H. Richter[a,c], P. Weigell[a]

*a Max-Planck-Institut für Physik, Föhringer Ring 6, D-80805, Munich, Germany*
*b CERN-PH, Switzerland*
*c Max-Planck-Institut Halbleiterlabor, Otto-Hahn Ring 6, D-81739, Munich, Germany*
*d now at DPNC, Université de Geneve (CH))*

**Abstract**

We present the results of the characterization of novel n-in-p planar pixel detectors, designed for the future upgrades of the ATLAS pixel system. N-in-p silicon devices are a promising candidate to replace the n-in-n sensors thanks to their radiation hardness and cost effectiveness, that allow for enlarging the area instrumented with pixel detectors.
The n-in-p modules presented here are composed of pixel sensors produced by CiS connected by bump-bonding to the ATLAS readout chip FE-I3. The characterization of these devices has been performed with the ATLAS pixel read-out systems, TurboDAQ and USBPIX, before and after irradiation with 25 MeV protons and neutrons up to a fluence of $5 \times 10^{15}$ $n_{eq}/cm^2$. The charge collection measurements carried out with radioactive sources have proven the feasibility of employing this kind of detectors up to these particle fluences. The collected charge has been measured to be for any fluence in excess of twice the value of the FE-I3 threshold, tuned to 3200 e.
The first results from beam test data with 120 GeV pions at the CERN-SPS are also presented, demonstrating a high tracking efficiency before irradiation and a high collected charge for a device irradiated at $10^{15}$ $n_{eq}/cm^2$.
This work has been performed within the framework of the RD50 Collaboration.

---

* *E-mail address*: annamac@mpp.mpg.de.

**1. Introduction**

The upgrade to the Large Hadron Collider (LHC), known as the High Luminosity-LHC or HL-LHC, is planned to be achieved in two steps [1]. From 2016 the luminosity is foreseen to be increased up to around $(2\text{-}3)\times 10^{34}$ cm$^{-2}$s$^{-1}$. An upgrade of the pixel system, called NewPix, is under discussion for the year 2018, where, depending on the operational performance, a replacement of the entire pixel detector may be needed. After 2020, a major upgrade will increase the luminosity further to $5\times 10^{34}$ cm$^{-2}$s$^{-1}$, that will require the replacement of the ATLAS strip tracker and of the pixel system, in case the latter will not have been already exchanged in 2018. In this scenario the innermost layers of the ATLAS vertex detector system will have to sustain very high integrated fluences of more than $10^{16}$ n$_{eq}$/cm$^2$ (1MeV neutron equivalent). The total surface covered by pixel sensors is foreseen to increase from the present 1.8 m$^2$ to about 10 m$^2$ in the upgraded ATLAS detector. This significantly larger area requires the use of cost-effective n-in-p sensors, requiring only a single-sided process, instead of the standard n-in-n technology used to instrument the pixel detectors of the LHC experiments.

**2. N-in-p pixel sensor production and high voltage stability**

A production of n-in-p pixel sensors, with a geometry compatible with the ATLAS FE-I3 chip, was realized at CiS [2] in the framework of the RD50 Collaboration. The detectors were processed on 4" diffusion oxygenated FZ wafers, 285 µm thick, with a bulk resistivity of 10 kΩcm.

The n-in-p pixel sensors require a single-sided process, leading to a cost reduction with respect to the double sided process of the n-in-n pixel devices. Thanks to the absence of bulk type inversion and to the fact that the main junction is always between the n$^+$ implanted pixels and the p-type bulk, the guard ring structure can be placed on the front-side (see Fig.1) while the back-side is implanted with a uniform p$^+$ implantation.

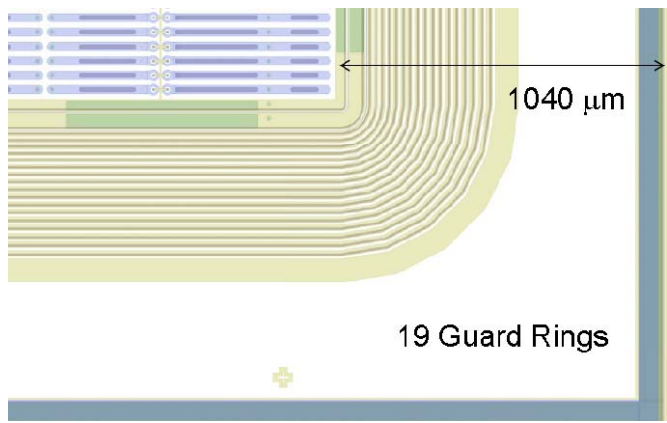

Fig.1 Design of the guard-ring structure with 19 rings on the front side of the n-in-p pixel sensors. The back-side of the sensor (not shown in the picture) is uniformly implanted with boron.

Two p-spray methods have been implemented in the wafers: moderated p-spray, as in the n-in-n pixel sensors presently instrumenting the ATLAS detector, and uniform p-spray, that has shown comparable or slightly better performance in previous n-in-p pixel sensor productions [3]. The dose of the moderated p-spray is modulated through the nitride layer while the homogenous p-spray is uniformly implanted into the wafer. The IV-curves of the FE-I3 pixel sensors in Fig.2 show that the breakdown voltages are well in excess of the depletion voltage of 60V before irradiation and the leakage currents are not affected by the bump bonding interconnection procedure.

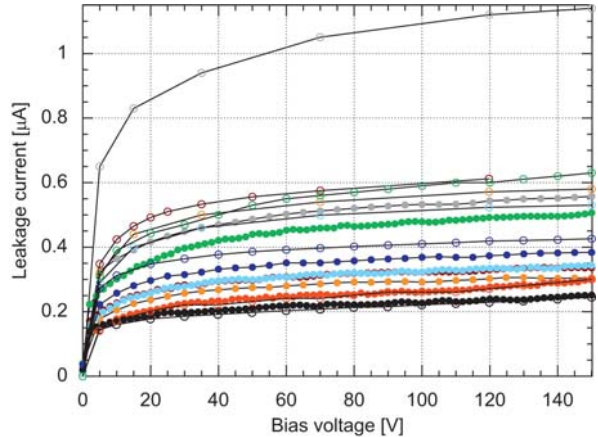

Fig.2 IV characteristics of FE-I3 n-in-p sensors, before (filled symbols) and after (open symbols) interconnection.

In comparison with the n-in-n devices, the n-in-p pixel modules pose the additional challenge of the isolation between the sensor, with the edges outside the guard rings lying at the high potential of the backside, and the chip that is at ground potential. A layer of 3 µm of Benzocyclobutene (BCB) has been applied to the sensor surface at the wafer level, in addition to the original $SiO_2$ passivation, to improve on the electrical isolation to the chip.

**3. Characterization of the n-in-p modules before irradiation**

The interconnection of sensors and FE-I3 chips has been performed by IZM-Berlin using the bump-bonding technique. The modules have been fully characterized before irradiation [4] using the TurboDAQ and USBPix systems [5], developed and used within the ATLAS pixel detector community to test the FE-I3 devices. The FE-I3 chip can be optimized for different working conditions with a set of internal configuration parameters, controlled by digital to analog converters (DAC). The tuning procedure allows for obtaining a homogeneous threshold setting among different pixels, constraining response differences, to the achievable threshold dispersion. When a charge deposited in the sensor is above the discriminator threshold, the front-end electronics stores the Time-over-Threshold (ToT), i.e. the time during which the pre-amplifier output is above threshold. By design the ToT has a nearly linear dependence on the amplitude of the pre-amplifier output and therefore on the charge released in the sensor.

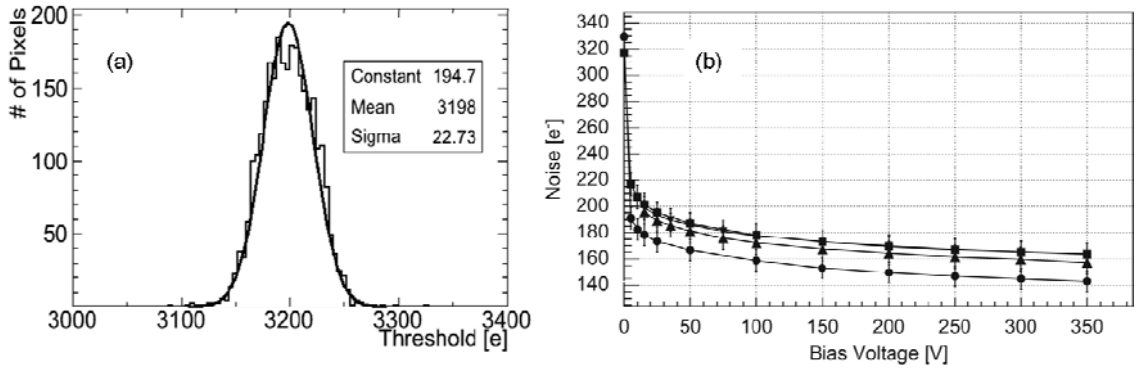

Fig 3. (a) Threshold distribution for a FE-I3 n-in-p module before irradiation, after tuning to a target threshold value of 3200 e. (b) Threshold noise dependence on the applied bias voltage, measured on a sub-sample of the not-irradiated modules.

Very uniform results have been obtained among the 19 modules that have been produced. Before irradiation the chips can be tuned to values of the discriminator thresholds between 2500 and 3200 electrons. A typical example of the threshold distribution is shown in Fig.3a, with a dispersion of 23 electrons. A systematic study on a sub-sample of the not irradiated modules showed that the threshold noise decreases with bias voltage (see Fig.3b), following the inter-pixel capacitance behavior. At a bias voltage of 150V the threshold noise for the different modules lie in the range from 140 to180 e, comparable to what has been achieved for the n-in-n modules.

Charge Collection Efficiency (CCE) measurements have been performed with $^{90}$Sr and $^{241}$Am radioactive sources [5]. In case of $^{90}$Sr scans, a charge of 19-20 ke is measured before irradiation as Most Probable Value (MPV) of the Landau distribution (Fig.4a). The most relevant source of systematic uncertainties is due to the ToT to charge calibration procedure, and it accounts for an uncertainty of 10% on the charge. The dependence of the MPV on the applied bias voltage is illustrated in Fig. 4b and it shows an agreement with the depletion voltage of 60V as measured on the bare sensors.

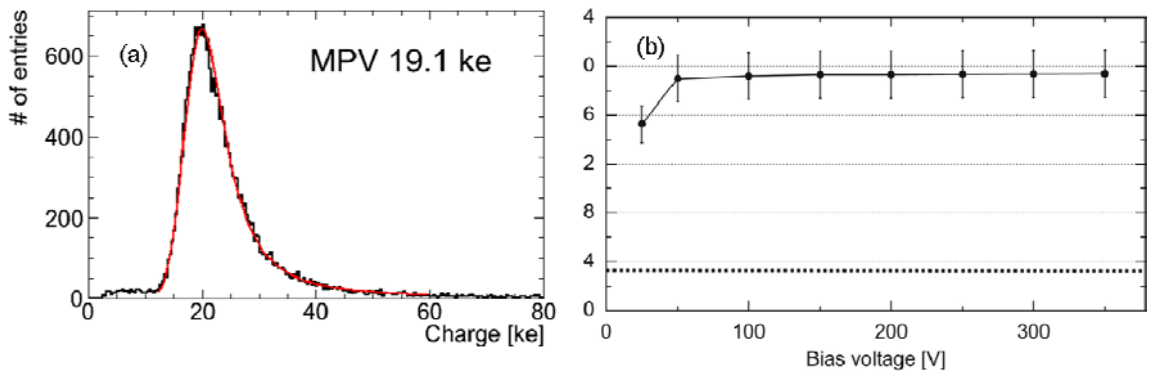

Fig. 4. (a) Distribution of the charge collected with a $^{90}$Sr source scan, in a not-irradiated n-in-p module, biased at a voltage of 150 V. (b) Dependence of the MPV on the applied bias voltage, obtained with $^{90}$Sr scans, in a not-irradiated module.

## 4. Characterization of the modules after irradiation

An extensive irradiation program has been completed, with 25 MeV protons at the Karlsruhe Institute of Technology (KIT), at a fluence of $10^{15}$ $n_{eq}/cm^2$ and with reactor neutrons in the Josef-Stefan-Institut in Ljubljana up to a fluence of $5 \times 10^{15} n_{eq}/cm^2$ [6]. The performance of the proton irradiated detectors at $10^{15}$ $n_{eq}/cm^2$ is reported in [5] while this paper is focused on the results obtained with the neutron irradiated samples. The neutron irradiated modules were characterized with $^{90}$Sr scans after tuning the chip to a threshold of 3200 electrons. The threshold noise, up to the highest fluence of $5 \times 10^{15}$ $n_{eq}/cm^2$, being comprised in the range between 150 and 200 e, is still comparable with pre-irradiation values. The threshold distribution is still very narrow for bias voltages below 800V and only for the highest voltages, up to the maximum value of 1000V, the dispersion increases. As shown in Fig.5a, even in this case the threshold is always an order of magnitude larger than the noise (Fig.5b), a condition needed to ensure an acceptable level of noise occupancy.

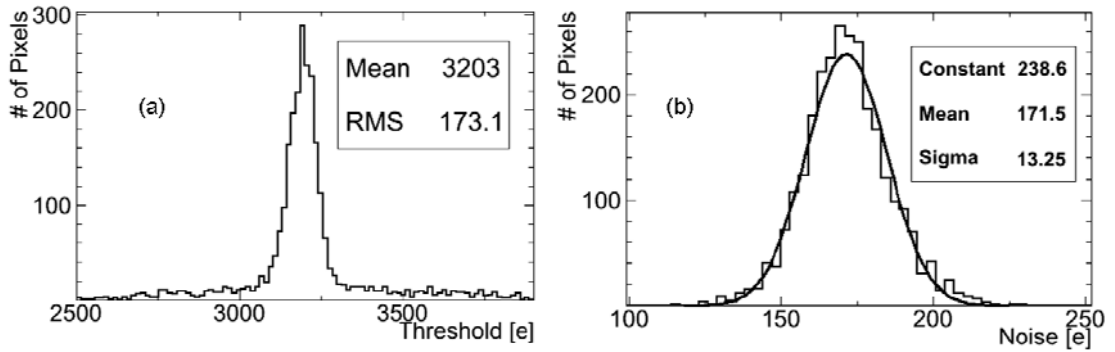

Fig. 5. Threshold (a) and noise distribution (b) for a n-in-p module irradiated at a fluence of $5 \times 10^{15} n_{eq}/cm^2$, tuned to a threshold value of 3200 e and biased at 1000V.

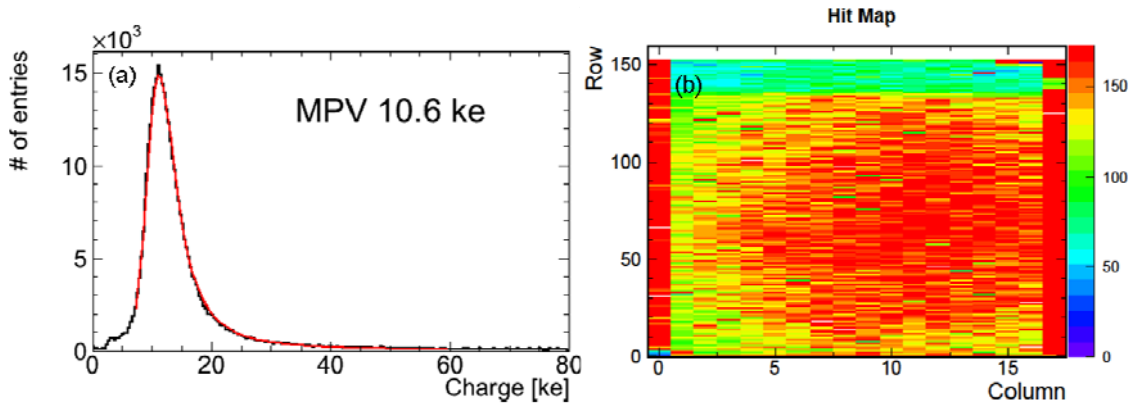

Fig. 6: (a) Landau distribution and (b) hit map resulting from a $^{90}$Sr source scan for a n-in-p module irradiated at a fluence of $5 \times 10^{15} n_{eq}/cm^2$, biased at 1000V. The top seven rows, populated with ganged and inter-ganged pixels, are masked.

The Landau distribution (Fig.6a) of the device irradiated to $5 \times 10^{15}$ $n_{eq}/cm^2$ in the $^{90}$Sr source scan at 1000V is characterized by a MPV of 10.0 ke. The corresponding hit map (Fig. 6b) indicates that all channels of the module are connected and functional. The upper seven rows, populated with ganged and inter-ganged pixels and normally characterized by a higher noise [4], were not included in the analysis,

A summary of the collected charge obtained after neutron irradiations is given in Fig.7, for a tuning to a discriminator threshold of 3200 electrons: the MPV at equal applied voltages decreases with the received fluence, but all samples can deliver a charge well above twice the threshold. These results are compared with the measurements obtained with an n-in-n detector, irradiated to $5 \times 10^{15} n_{eq}/cm^2$, also connected to the FE-I3 chip [7]. Although the n-in-n module delivers slightly higher charge for bias voltages lower than 800V, above this voltage the two technologies offer very comparable performance.

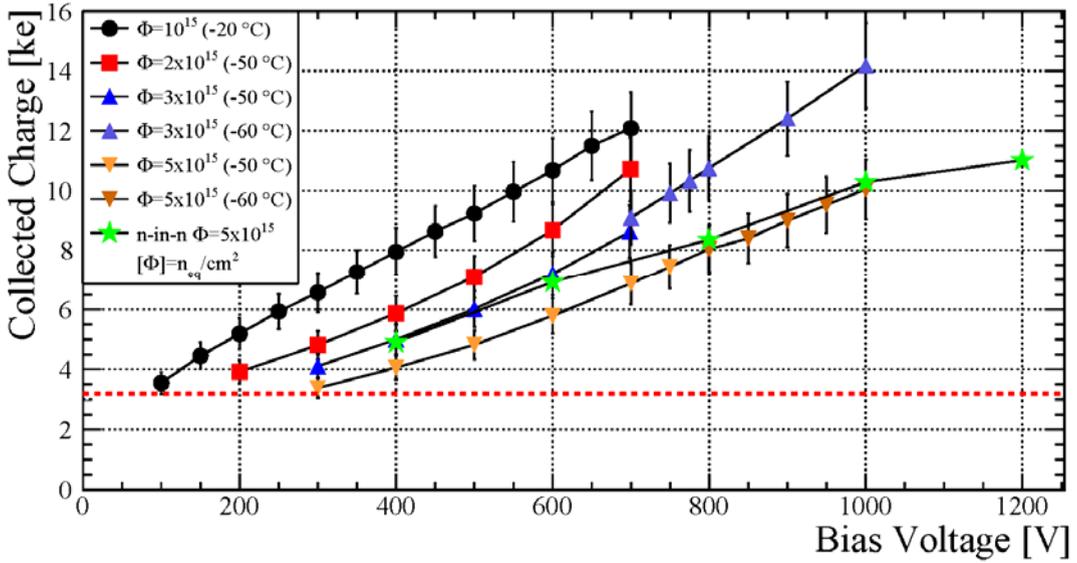

Fig. 7. Charge collected by neutron irradiated n-in-p modules in $^{90}$Sr scans after tuning to a discriminator threshold of 3200 e. Data for the n-in-n module are extracted from [7]. The temperature specified in the plot is relative to the environmental conditions of the climate chamber in which the measurements were performed and not to the temperature of the module itself.

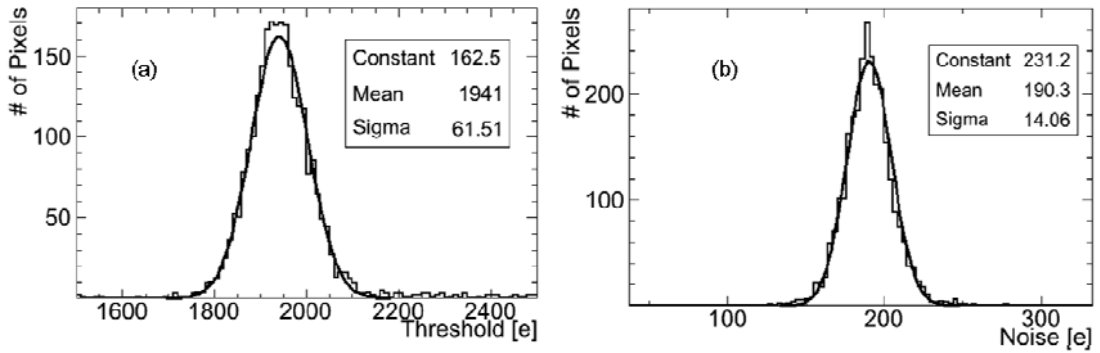

Fig. 8. Threshold and noise distribution for the sample irradiated at $5 \times 10^{15} n_{eq}/cm^2$, biased at 600V and tuned to a target threshold of 2000 e.

The target threshold of 3200 e was chosen to obtain a consistent set of results with the beam-tests, where these devices were also studied. A tuning of the FE-I3 chip to lower threshold values was performed to operate these devices at lower bias voltages, while maintaining a similar charge to threshold ratio. The module irradiated to $5\times10^{15}$ $n_{eq}/cm^2$ has been tuned to a threshold of 2000e (Fig. 8a) and studied with a $^{90}$Sr scan at 600V. The MPV of the collected charge is 5.5 Ke, above twice the threshold, while the noise (Fig.8b) is measured to be 190 e. The noise occupancy has been estimated to be $10^{-6}$ per pixel, by performing a run without the source and using the internal trigger capability of the FE-I3 chip. These results suggest the possibility of operating n-in-p modules, irradiated to $5\times10^{15}$ $n_{eq}/cm^2$, at a bias voltage of 600V, although the noise occupancy measurement needs to be confirmed in the upcoming beam tests.

First results with the new ATLAS front-end chip FE-I4, developed for the IBL upgrade, indicates that thresholds as low as 1.6 ke are achievable [8], which would allow to operate pixel detectors irradiated up to $5\times10^{15}$ $n_{eq}/cm^2$ in a moderate bias voltage range (<500V).

The irradiated n-in-p modules have been operated during the source scans for several hours at a bias voltage of 1000V and for a few days at 800V during the beam tests in both cases without sparks occurring at the edges between the sensor and the chip. The adopted solution of the BCB passivation, proven to work for this limited sample of modules, will be compared in the future with alternative techniques, like for example a parylene coating.

## 5. Test-beam results

N-in-p modules of this production were tested in 2010 at CERN SPS/H6 beam-line with 120 GeV pions using the EUDET telescope [9,10]. The results of the not-irradiated module, providing a mean tracking efficiency of 99.3%, were reported in [5].

In a later test-beam a module irradiated to $10^{15}$ $n_{eq}/cm^2$ was studied. Due to a problem in the synchronization between the telescope planes and the devices under test, no tracking efficiency results are available at the moment for this irradiated module. The sub-pixel tracking resolution of the beam telescope allows for building a map of the charge collection within a pixel, obtained by superimposing all pixels of the sensors. Fig. 9a shows the average charge deposit at a bias voltage of 500V, where the cluster position is displayed with respect to the reconstructed impact point of the pion. The regions of lower charge collection efficiency correspond to the four corners, affected by a higher charge sharing and to the bias dot (Fig. 9b). Nonetheless also these areas can deliver a charge that is well above twice the threshold of 3200 e.

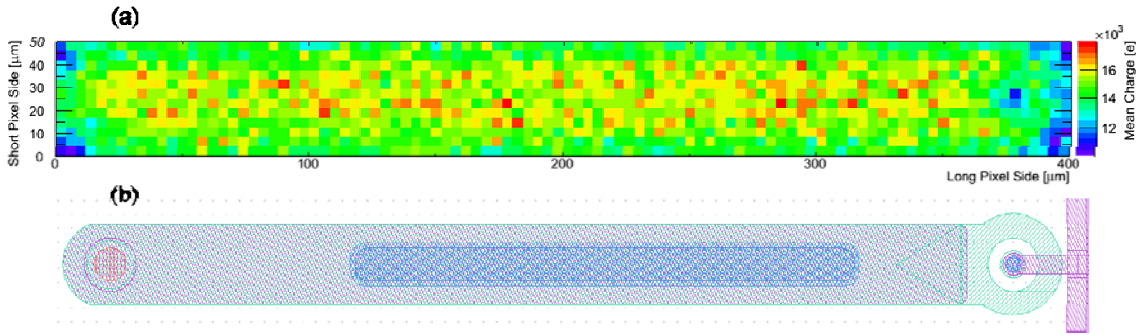

Fig. 9. (a) Map of the mean collected charge at a bias voltage of 500V, for a n-in-p module irradiated to $10^{15}$ $n_{eq}/cm^2$ obtained superimposing all the pixels of the sensor. The region of lower efficiency on the right side (circled) correspond to the position of the bias dot (b).

## 6. Conclusions

Pixel modules, composed of n-in-p sensors and the ATLAS FE-I3 chip, have been characterized before and after irradiation up to $5 \times 10^{15}$ $n_{eq}/cm^2$. Good performances in terms of signal over threshold and noise have been obtained: at the maximum fluence about 10 ke electrons were collected at a bias voltage of 1000V with the discriminator threshold tuned to 3200 e. The results reported in the paper also suggest the possibility of operating n-in-p modules, irradiated to $5 \times 10^{15}$ $n_{eq}/cm^2$, at a bias voltage of 600V. No sparks between chip and sensor were observed up to 1000V thanks to the coating of the sensor surface with a Benzocyclobutene layer.

## Acknowledgements

We are grateful to I. Mandic for the irradiation with reactor neutrons in the Josef-Stefan-Institut in Ljubljana, supported by the AIDA package WP7, and to A. Dierlamm, for the irradiation at KIT, supported by the Initiative and Networking Fund of the Helmholtz Association, contract HA-101 ("Physics at the Terascale"). This work has been carried out within the RD50 Collaboration.